\documentclass[usenatbib,usegraphicx,useAMS]{mn2e}

\usepackage{times}

\newcommand{\fig}{Fig.\,}
\newcommand{\figs}{Fig.'s~}

\newcommand{\tabl}{Table\,}

\newcommand{\kmsmpc}{\mbox{km\,s$^{-1}$\,Mpc$^{-1}$}}
\newcommand{\magn}{\mbox{$^{\mathrm m}$}}

\begin{document}
\title[PLE: Missing Bright Galaxies at High Redshift]{Pure Luminosity
  Evolution Models: Too Few Massive Galaxies at Intermediate and High 
Redshift}
\author[Kitzbichler et al.]{M.~G.~Kitzbichler \and S.~D.~M.~White\\
Max-Planck Institut f\"ur Astrophysik, D-85748 Garching, Germany}
\maketitle
\begin{abstract}
  We compare pure luminosity evolution (PLE) models with recent data at low
  and high redshift. These models assume that massive galaxies were assembled
  and formed most of their stars at high redshift ($z>3$) and have evolved
  without merging or substantial dust obscuration since then. Our models span
  the full range of plausible metallicities, initial mass functions (IMF's)
  and star formation histories. We require them to reproduce the abundance of
  galaxies by colour and luminosity in the Sloan Digital Sky Survey and we
  investigate whether they can simultaneously fit (i) the observed galaxy
  counts as a function of redshift in magnitude limited surveys with $K<20$,
  and (ii) the colour and $M/L$ ratio evolution of red sequence galaxies in
  clusters. All models that are consistent with (ii) predict galaxy counts at
  $1.5<z<3$ which lie above the observations. Models with an IMF slope similar
  to the Salpeter value lie far above the data.  We conclude that the majority
  of massive galaxies were either assembled relatively late in this redshift
  interval or were substantially obscured by dust at these redshifts.
\end{abstract}
\nokeywords

\section{Introduction}
\label{sec:intro}
A very wide range of possible evolutionary histories appear consistent with
the observed properties of the present-day population of galaxies.  The
simplest and most conservative assumption may be that most galaxies were
assembled at some early time and their differing stellar populations reflect
differing subsequent star formation histories.  Massive galaxies -- big
ellipticals, S0's and early-type spirals -- appear to be dominated by old
stellar populations, so their star formation rates (SFR) must have been high
at early times and must thereafter have declined steeply. Many less massive
galaxies -- late-type spirals and irregulars -- show evidence for substantial
recent star formation, so their SFR's may have varied much less. The light of
some is clearly dominated by stars from a recent burst.

In order to model recent evolution of the galaxy population in such a scenario
one can adopt the backwards-in-time technique first introduced by Tinsley
\citep[see][]{1980ApJ...241...41T}. This requires three main ingredients: the
present-day luminosity function (LF) of galaxies divided by morphological type
(or better by colour); a parametrisation of the mean star formation history
(SFH) for each type (or colour class); and a global cosmological model to
relate times, distances and redshifts. The SFH is fed into stellar population
synthesis models which determine how the luminosities and colours of
each type evolve with time. These can then be combined with the cosmological
model to predict counts of galaxies as a function of apparent magnitude,
observed colour and redshift.

\citet[][ KC98 hereafter]{1998MNRAS.297L..23K} compared available data to the
redshift distribution predicted for complete $K$-band-limited galaxy samples
by such pure luminosity evolution (PLE) models assuming an Einstein--de Sitter
cosmology. They found the models to overpredict counts at redshifts $z>1$ by a
large factor. Since then a number of similar studies have updated the
cosmological model to the current concordance cosmology and have presented new
observational samples which cover wider areas or go significantly
deeper. While the improved observations have significantly reduced the
statistical uncertainties, they have not substantially changed the redshift
distributions from those used by KC98. The change to $\Lambda$CDM
significantly reduced the discrepancy, however, by bringing down the number of
high redshift objects predicted at a given K magnitude.

\citet{1999MNRAS.310L..27F} published a study based on photometric
redshifts for a $K\le 21$ sample of 319 galaxies in several small
fields. Despite using a $\Lambda$CDM model their conclusion agreed
with KC98; the observed redshift distribution disagreed with their PLE
model.  \citet{2001AJ....122.2205R} found the same result when
comparing a range of published PLE models with their photometric
redshifts for 95 $K_{s,AB}\le 22$ galaxies in the Hubble Deep Field
South.  In part II of a series of papers on the Las Campanas Infrared
(LCIR) Survey \citet{2002MNRAS.332..617F} present photometric
redshifts for 3177 galaxies down to $H\le 20$. They compare these to a
number of different PLE models and again find the abundance of high
redshift objects to be overpredicted.  All these studies echoed the
KC98 conclusion that the data suggest that many present-day massive
galaxies were assembled at relatively low redshift.

Other recent work based on similar data disagrees with this conclusion.
\citet{2003AJ....125...53K} and \citet{2002A&A...391L...1C} both compare to a
modified ``PLE'' model by \citet{2001ApJ...559..592T} which incorporates dust
and high-z selection effects, as well as a simplified parametrisation of
mergers. This model is able to fit the observed redshift distributions because
its large assumed dust extinction hides most massive galaxies at redshifts
beyond 1.5 or so. In this article we are primarily concerned with traditional
PLE models in which mergers are neglected and extinction is assumed weak, in
particular for massive galaxies after their initial burst of star
formation. We will, however, comment briefly on the effects of dust.

One of the most recent studies comparing PLE predictions to the redshift
distributions of K-selected samples is that of \citet{2004ApJ...600L.135S} who
found that although such models overproduce the counts at high redshift, the
discrepancy is quite modest. They took advantage of the newly acquired K20 and
GOODS survey data, which we also use here, together with other recent high
quality survey data, for comparison to our own PLE models.  As we will see,
our conclusions do not agree with those of \citet{2004ApJ...600L.135S} even
for similar models.

In this letter we investigate a number of traditional PLE models spanning the
full plausible range of metallicity, initial mass function (IMF), and star
formation history. The following Section \ref{sec:models} describes how our
models are set up to reproduce the present-day LF's as a function of colour in
the Sloan Digital Sky Survey (SDSS) (\S\ref{subsec:locallf}) and how various
different SFR's and metallicities are assigned to the different colour
classes (\S\ref{subsec:modelconst}) in order to follow their luminosity
evolution backward in time. We establish the range of allowed parameters and
present five models to illustrate the resulting range of evolutionary
predictions. We check that our models reproduce the local K-band LF, as
observed by the 2MASS survey (\S\ref{subsec:kbandlf}) as well as the passive
evolution of colour and M/L ratio observed for cluster elliptical galaxies. In
Section \ref{sec:results} we compare the predictions of these models with
counts as a function of redshift in recent deep K-selected surveys.  Finally
in the concluding Section \ref{sec:discussion} we discuss possible
interpretations of our primary result, that there are fewer luminous galaxies
observed at $z\ga 1.2$ than are expected on the basis of traditional PLE
models. Either a large amount of dust obscures galaxies at higher redshifts, or
many present-day massive galaxies were not yet assembled by $z\sim 2$.

\section{The Models}
\label{sec:models}
As mentioned above, traditional PLE models require knowledge of the
present-day LF's of galaxies as a function of their colour.  For each colour
class a star formation history (SFH) model is assumed which reproduces its
$z=0$ colour, and this SFH is then used to predict the LF and the spectral
energy distribution (SED) of galaxies of this class at all earlier
times. Combining the different classes, galaxy counts can then be predicted as
a function of observed magnitude, colour and redshift in any observed
photometric band for any assumed cosmological model. In the following we adopt
the cosmological parameters of the present standard concordance
cosmology: $\Omega_\rmn{M}=0.3$, $\Omega_\Lambda=0.7$, and $H_0=70~\kmsmpc$.

\subsection{From the local LF to the models}
\label{subsec:modelconst}
\begin{figure}
\begin{center}
\includegraphics[width=\linewidth]{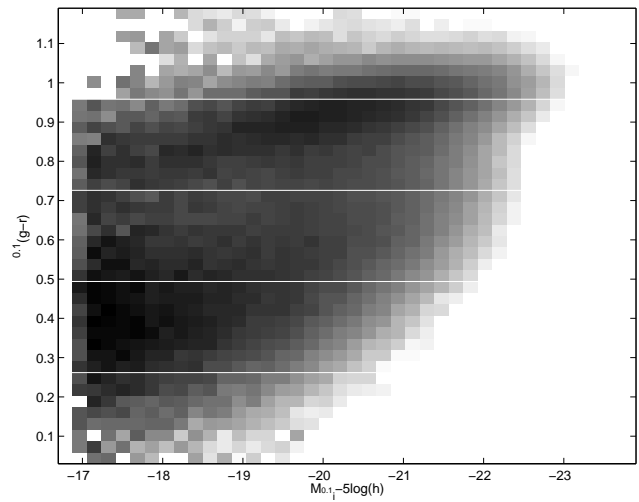}
\caption{\label{condlf}Two dimensional luminosity functions by colour and
  absolute magnitude taken from \citet{2003ApJ...594..186B}. White lines
  indicate the colours separating our different colour classes.}
\end{center}
\end{figure}

\begin{figure}
\begin{center}
\includegraphics[width=\linewidth]{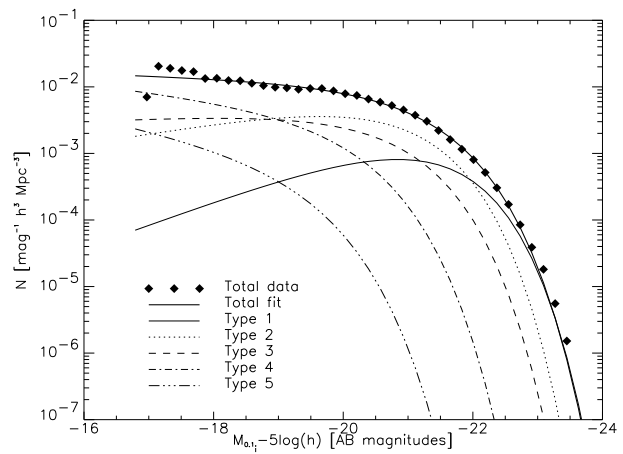}
\caption{\label{condlfparm}Schechter function fits to the luminosity functions
  of SDSS galaxies in our five different colour classes (see also
  \tabl\ref{lfdeftab}).}
\end{center}
\end{figure}

%\subsection{The local luminosity function}
\label{subsec:locallf}
Our PLE models are normalised to the luminosity functions at redshift
$z_\rmn{LF}=0.1$ recently obtained by \citet{2003ApJ...594..186B} from the
data of the SDSS survey \citep{2000AJ....120.1579Y}. For our purposes the
great advantages of these data are their high quality, their superb
statistical precision and the fact that they are given in colour-luminosity
space (see \fig\ref{condlf}). We separate the data distribution into five
colour ranges and calculate the parameters (see \tabl\ref{lfdeftab}) for a
Schechter function fit to the LF of each colour bin independently.  These
parametrised LF's are shown in \fig\ref{condlfparm}.

\begin{table}
\caption{Definition of the different galaxy types according to their
  colour. Also the parameters of the Schechter function fits to the
  respective LF's are given here.\label{lfdeftab}}
\begin{center}
\begin{tabular}{ccccc@{\hspace{2mm}}c@{\hspace{2mm}}c}
\hline\hline
 & \multicolumn{ 2}{c}{\scshape Colour} $^{0.1}(g-r)$ & &
 \multicolumn{ 3}{c}{\scshape LF -- Schechter fit}\\
 \cline{5-7}\cline{2-3}
{\raisebox{1.5ex}[-1.5ex]{\scshape Type}} & {\scshape mean} & {\scshape range} & &
$\Phi[\mbox{Mpc/h}]^{-3}$ & $\alpha$ &
$_{^{0.1}i}M_*$\\
\hline
1\dotfill & 1.01 & $0.96\dotfill 1.19$ & & $2.377\ 10^{-3}$ & $-0.11$ & $-20.96$ \\
2\dotfill & 0.87 & $0.73\dotfill 0.96$ & & $8.406\ 10^{-3}$ & $-0.60$ & $-20.61$ \\
3\dotfill & 0.61 & $0.49\dotfill 0.73$ & & $5.169\ 10^{-3}$ & $-0.89$ & $-20.49$ \\
4\dotfill & 0.40 & $0.26\dotfill 0.49$ & & $4.382\ 10^{-3}$ & $-1.29$ & $-19.84$ \\
5\dotfill & 0.20 & $0.03\dotfill 0.26$ & & $9.596\ 10^{-4}$ & $-1.51$ & $-19.11$ \\
\hline
\end{tabular}
\end{center}
\end{table}

%\subsection{Constructing the models}
%\label{subsec:modelconst}
We use the fits of \fig\ref{condlfparm} to construct PLE models as described
in \citet{1998PASP..110..291G} -- except for the slight complication
that $z_\rmn{LF}=0.1$. The five colour classes are identified with five SFH's
which reproduce their broad-band colours according to the stellar population
synthesis models of \citet{2003MNRAS.344.1000B}. For each galaxy type the
spectrum and the LF can then be evolved backwards in time in order to predict
the properties of the galaxy population at earlier redshifts.

The assignment of SFH to present-day colour is far from unique, so we
construct a variety of possible models differing in their IMF, metallicity,
formation redshift $z_\rmn{f}$ (defined as the redshift when stars start to
form) and e-folding timescale $\tau$ for an assumed exponentially declining
SFR. We assume all colour classes to have the same $z_\rmn{f}$, except for the
bluest one, which often cannot be fit by any exponentially declining SFR. This
is a particular problem for models with a steep IMF. In such cases we assume a
SFH with constant SFR seen at a fixed age, implying {\it no} evolution with
redshift.  This is the standard fix for this problem, which is, in any case,
irrelevant for the questions we study here.

\begin{figure}
\begin{center}
\includegraphics[width=\linewidth]{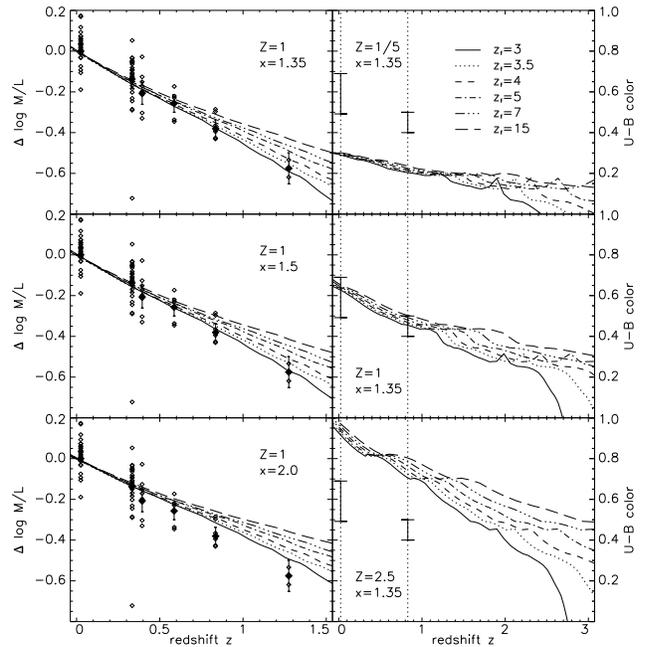}
\caption{{\em Left:} Evolution of the mass-to-light ratio of cluster
  ellipticals in the $B$-band as given by \citet{2003ApJ...585...78V}. Small
  open symbols denote individual galaxies while big filled symbols stand for
  data averaged over a number of massive galaxies in a cluster. The model
  predictions are shown for different $z_\rmn{f}$ and IMF slopes ranging from
  $x=1.35$ at the top to $x=2$ at the bottom.  {\em Right:} Rest-frame
  $U$-$B$ evolution of model early-type galaxies compared to the rest-frame
  $U$-$B$ colours of cluster ellipticals at $z=0.87$ (MS 1054-03) and at
  $z=0.023$ (Coma).  Model predictions are shown for different $z_\rmn{f}$ and
  for three metallicities, $0.2Z_\odot$, $Z_\odot$ and $2.5 Z_\odot$ from top
  to bottom.\label{mlrubcol}}
\end{center}
\end{figure}

We limit the range of allowed parameters in our PLE models by requiring
consistency with the observed, apparently passive evolution of bright
early-type galaxies in clusters. We require the $B$-band mass-to-light ratio
of our reddest colour classes to evolve similarly to the measurements of
\citet{2003ApJ...585...78V}.  As the left three panels in \fig\ref{mlrubcol}
show, this mainly constrains the slope of the IMF, given that one has
considerable freedom in the choice of the formation redshift
$z_\rmn{f}$. IMF's with a power law exponent of $x=2.0$ (where the
Salpeter exponent is $x=1.35$) are excluded, except possibly for the lowest
formation redshifts.  We nevertheless adopt this slope for Model~4 below in
order to study its implications. We note that most recent work on IMF's at
high redshift have tended to argue for exponents significantly flatter than
Salpeter (``top-heavy IMF's'') in order to explain the high luminosities of
sub-millimeter luminous galaxies and the apparently high aggregate metal
yields of early generations of stars.

We also require the rest-frame $U$-$B$ colours of the reddest colour
class to match those of bright ellipticals in two clusters, the Coma
cluster at $z=0.023$ and MS 1054-03 at $z=0.87$
\citep{1991AJ....101.1207G,1999ApJ...520L..95V}. This allows only a
narrow range of metallicities for these bright early-types, namely
approximately solar, as can be seen from the three right-hand panels
in \fig\ref{mlrubcol} which show the evolution in rest-frame colour
for stellar populations of given metallicity formed with a Salpeter
IMF in a single burst at a variety of redshifts. IMF variations have
very little effect on this colour since it is dominated by main
sequence turn-off stars \citep[as explained
by][]{2003MNRAS.344.1000B}.

We present results for five representative models that are at least marginally
consistent with all these constraints.  Their parameters are summarised in
\tabl\ref{moddeftab} and were selected to cover the whole range of permitted
values.

\begin{table}
\caption{Definition of the different models. The given parameters are:
  slope of the IMF -- x, formation redshift -- $z_\rmn{form}$, and
  exponential fall-off time of the SFR -- $\tau$ (where $\infty$ means
  constant star formation).\label{moddeftab}}

\begin{tabular*}{\linewidth}{l@{\extracolsep{\fill}}rrrrr}
\hline\hline
{\scshape Model}         & 0    & 1    & 2    & 3    & 4\\
\hline
{\scshape imf} x         & 1.35 & 1.5  & 1.35 & 1.5  & 2.0\\
$z_\rmn{form}$           & 15   & 15   & 3.5  & 3.5  & 3.5\\
\hline
$\tau_1$\dotfill         & 1.5  & 2.0  & 1.5  & 1.5  & 1.5\\
$\tau_2$\dotfill         & 3.0  & 3.0  & 2.5  & 2.5  & 3.0\\
$\tau_3$\dotfill         & 6.0  & 10.0 & 5.0  & 7.0  & 30.0\\
$\tau_4$\dotfill         &$\infty$&$^*\infty$&$\infty$&$^*\infty$&$^*\infty$\\
$\tau_5$\dotfill         &$^*\infty$&$^*\infty$&$^*\infty$&$^*\infty$&$^*\infty$\\
\hline
\end{tabular*}\\[0.5ex]
The $^*$ denotes galaxy types without evolution.\hfill
\end{table}

\subsection{The $K$-band LF as a consistency test}
\label{subsec:kbandlf}
\begin{figure}
\begin{center}
\includegraphics[width=\linewidth]{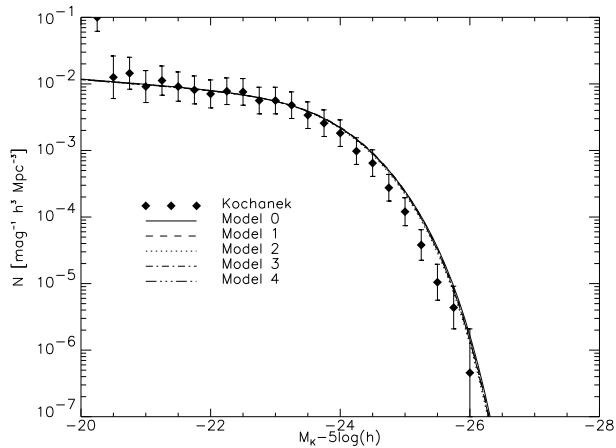}
\caption{Comparison of our model $K$-band LF's with the one that
  \citet{2001ApJ...560..566K} derived from 2MASS data. The slight offset at
  bright magnitudes can be accounted for by differing magnitude definitions in
  the SDSS and 2MASS surveys.\label{kbandlf}}
\end{center}
\end{figure}

The LF's used here were measured in the rest frame $^{0.1}i$-band. We
can check the reliability of our stellar population models for the
five colour classes by using them to predict the $K$-band
($2.2\,\umu\rmn{m}$) luminosity function of local galaxies. This is of
particular interest because near-IR light is a relatively good tracer
of stellar mass, depending only weakly on dust content and SFH. We
therefore compare the present-day $K$-band LF produced by our models
to the observed function as given by \citet{2001ApJ...560..566K}. As
can be seen in \fig\ref{kbandlf}, models and data agree reasonably
well apart from a slight magnitude offset, perhaps $\sim 0.15\magn$,
at the bright end.  This is likely due to the rather bright isophotal
magnitudes used by \citeauthor{2001ApJ...560..566K} in contrast to the
surface-brightness independent Petrosian magnitudes of the SDSS
survey. The difference is most pronounced for elliptical galaxies with
de Vaucouleur-type surface brightness profiles. These dominate the
bright end of the LF.

\section{Comparison of $K$-band Selected Redshift Distributions}
\label{sec:results}
\begin{figure}
\begin{center}
\includegraphics[width=\linewidth]{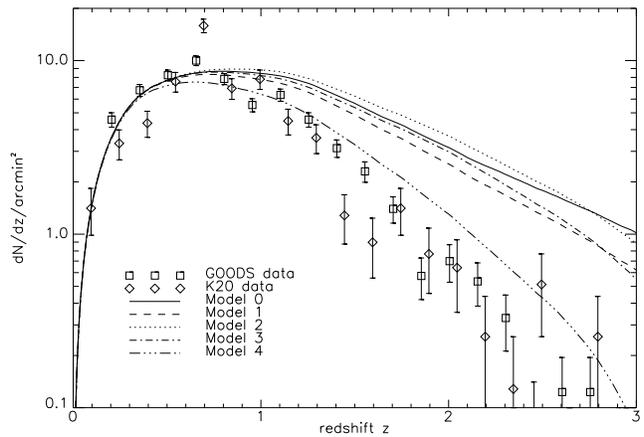}
\caption{Differential redshift distributions for $K<20$ galaxies.
  The errors plotted on the observational data points are approximate
  Poisson errors. Our 5 PLE models are shown as contiuous curves as
  indicated in the figure.\label{differential}}
\end{center}
\end{figure}

\begin{figure}
\begin{center}
\includegraphics[width=\linewidth]{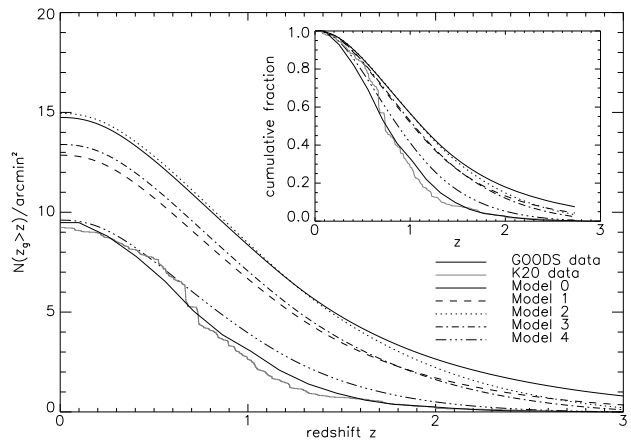}
\caption{Cumulative redshift distributions corresponding to the differential
  distributions of Fig.~5. The inset
  shows the same distributions normalised to $1.0$ at $z=0$ rather than the
  absolute counts per $\rmn{arcmin}^2$.\label{cumulative}}
\end{center}
\end{figure}

%\subsection{The observed redshift distribution}
%\label{subsec:zdistdata}
In this letter we compare to the same deep surveys as
\citet{2004ApJ...600L.135S}, namely GOODS CDF-S covering about
$160\,\rmn{arcmin}^2$ with photometric redshifts obtained by
\cite{2004ApJ...600L.167M} and K20 carried out in a smaller area of
the same field covering $52\,\rmn{arcmin}^2$ but providing
spectroscopic redshifts rather than photometric ones
\citep{2002A&A...392..395C}.  The differential distribution of
galaxies per $\rmn{arcmin}^2$ and per unit redshift interval is shown
in \fig\ref{differential} for both datasets, binned to $\Delta z=0.15$
and with Poisson errorbars. Clearly there is some substructure in
these distributions due to the relatively small fields surveyed. In
particular at $z\sim 0.7$ there is a prominent peak in the K20 data.
This feature is still visible in \fig\ref{cumulative}, the cumulative
redshift distribution of galaxies. In a larger comoving volume such
fluctuations should average out, which gives the smoother curves
obtained for the somewhat larger GOODS survey.

%\subsection{The predicted redshift distribution}
%\label{subsec:zdistpred}
Superposed on the observational data in \figs\ref{differential} and
\ref{cumulative} we show the differential and cumulative redshift
distributions predicted by the various models specified in
\tabl\ref{moddeftab}. In both figures directly predicted counts are
given per $\rmn{arcmin}^2$. In the inset of \fig\ref{cumulative},
however, we additionally show cumulative plots normalised to unity in
order to show that the predicted redshift distributions differ in
shape as well as in amplitude.

In order to quantify the obvious discrepancy between observations and
models, \tabl\ref{counttab} presents expected and measured counts
integrated over various redshift ranges. The standard Salpeter model,
model 0, overpredicts the observed counts beyond $z=1$ by a factor of
almost 3, beyond $z=1.5$ by more than a factor of 5, and beyond $z=2$
by nearly an order of magnitude.

\begin{table}
\caption{Predicted and observed galaxy counts in the range $1<z<3$.\label{counttab}}
\begin{center}
\begin{tabular}{*{6}{c}}
\hline\hline
 & \multicolumn{5}{c}{{\scshape Counts} [$\mbox{arcmin}^{-2}$]} \\
\cline{2-6}
{\raisebox{1.5ex}[-1.5ex]{\scshape Model}} & $z\ge1$ & $z\ge1.5$ & $z\ge2$ & $z\ge2.5$ & $z\ge3$ \\
\hline
0\dotfill & $ 8.36$ & $ 4.76$ & $ 2.65$ & $ 1.48$ & $ 0.80$ \\
1\dotfill & $ 6.66$ & $ 3.46$ & $ 1.70$ & $ 0.81$ & $ 0.35$ \\
2\dotfill & $ 8.47$ & $ 4.62$ & $ 2.22$ & $ 0.88$ & $ 0.21$ \\
3\dotfill & $ 7.05$ & $ 3.62$ & $ 1.62$ & $ 0.58$ & $ 0.11$ \\
4\dotfill & $ 3.96$ & $ 1.56$ & $ 0.50$ & $ 0.12$ & $ 0.01$ \\
\hline
K20\dotfill & $ 2.63$ & $ 0.73$ & $ 0.17$ & $ 0.06$ & $ 0.00$ \\
GOODS\dotfill & $ 3.04$ & $ 0.93$ & $ 0.29$ & $ 0.04$ & $ 0.00$ \\
\hline
\end{tabular}
\end{center}
\end{table}

\section{Discussion and Conclusions}
\label{sec:discussion}
The problem we study in this letter is whether the available
observational data are consistent with present day luminous galaxies
being already assembled with the bulk of their stars formed at high
redshift. If so, it should be possible to find a set of parameters
such that traditional PLE models can simultaneously reproduce: (i) the
present-day luminosity and colour distributions of massive galaxies;
(ii) the passive evolution in colour and M/L ratio observed for
massive early-type galaxies in clusters; and (iii) the observed galaxy
counts as a function of redshift in deep surveys.  Near-IR limited
surveys are best suited for this purpose since the observed magnitudes
are then a fair indicator of stellar mass and are only weakly affected
by dust. We therefore chose $K$-band data from the K20 and GOODS CDF-S
surveys for comparison with our models.

Out to redshift $z\sim1$ our model predictions are very similar to each other
and also fit the data reasonably, given their error bars.  At higher redshifts
all models predict too many galaxies. Only Model\,4, with $x=2$, comes close
to the data.  Obviously the IMF assumed has the largest impact on the
predicted number of galaxies at high redshift; the second and third best
models are the two with $x=1.5$. Changing the formation redshift only mildly
influences the shape of the distributions at $z<2.5$. The more conventional
standard Model\,0, using a Salpeter IMF, and its low $z_\rmn{f}$ pendant,
Model\,1, produce the predictions most inconsistent with the data. This may be
understood by recalling that the light of old stellar populations is dominated
by stars with masses near the main sequence turn-off. For younger populations
this turn-off is at higher masses. Hence a shallower IMF, corresponding to
more turn-off stars in younger populations, implies brighter galaxies at early
times, and so more high redshift galaxies above any apparent magnitude limit.
The $B$-band M/L ratio evolution of the brightest and reddest galaxies is an
important constraint on our models because it is also sensitive to the IMF for
the same reasons.  As already noted in Section\,\ref{subsec:modelconst} models
with $x=2$ are inconsistent with observation, except possibly for very low
formation redshifts.  Additionally, for $x=2$ our models cannot fit the
observed present-day colours of the bluest galaxy classes for constant SFR and
$z_\rmn{f}\ge 3.5$. For the galaxies to be blue enough they would have to be
much younger.  Missing bright blue stars at the high mass end of the IMF again
account for the effect. Finally, since most models for the light output and
metal production of high redshift galaxies require IMF's with substantially
{\it more} high mass stars than Salpeter \citep[e.g.][]{2004astro.ph..8529N}, an IMF as
steep as $x=2$ appears very unlikely as an explanation of the apparent lack
of high redshift massive galaxies.

Our Model\,0 is very similar to the PLE model used by
\citet{2004ApJ...600L.135S} but whereas we find it to be badly inconsistent
with the data, they conclude that any problem is marginal. There are two
reasons for this discrepancy. Looking at their Figure\,1 there is clearly a
problem in going from their differential redshift distribution, which is very
similar to our own, to the cumulative distribution, which predicts
substantially fewer high redshift galaxies than does ours. In addition, they
compare the cumulative distribution to the data after normalising both to
unity (as in the inset to \fig\ref{cumulative}) which then misses the fact
that the total predicted galaxy count at K$<20$ is substantially larger than
observed.

All of our models with $x\le 1.5$ overpredict the counts at redshifts $z>1$ by
a large factor as can be seen in \tabl\ref{counttab}. In the interval $1<z<2$
these models all predict more than twice the number of galaxies observed and
in the interval $2<z<3$ they are off by factors between 4 and 11. Could cosmic
variance or dust account for this? The clustering of galaxies has the greatest
effect at low redshift, where the observed volume is comparatively small and
clear evidence of large fluctuations is seen in \fig\ref{differential} at
$z=0.7$ in the K20 data. However in this range the models still agree quite
well with the data, only at higher redshifts do they deviate.  Also the model
predictions are obviously systematically too high at all $z$ which is not what
one would expect if the effect was due to cosmic variance. Finally, models and
data also disagree in the normalised version of the diagram (inset in
\fig\ref{cumulative}).

Extinction by dust, on the other hand, might indeed be important. As a
simple model to assess how much dust is required to bring our PLE
models into agreement with the data, consider placing a foreground
screen in front of all galaxies at $z = 1.5$, thereby translating
their apparent luminosity function fainter by some fixed amount. We
find that to lower the count for Model 0 in \fig\ref{differential} by
the factor of $2.1$ needed to bring it into agreement with the GOODS
data at this redshift requires $0.7$ magnitudes of extinction at
observed K (i.e. at rest-frame $z$). Carrying out a similar
calculation at $z=2$ we find that $1.0$ magnitudes of extinction is
again required at observed K (now rest-frame $r$) to reduce the
abundance of galaxies per unit redshift by the required factor of
$5.1$. For comparison \citet{2003MNRAS.341...33K} analysed dust
attenuation in a sample of 122808 low redshift galaxies drawn from the
SDSS, finding a typical (median) attenuation of 0.2 -- 0.3 magnitudes
in the $z$-band for massive galaxies. We thus need much more dust in
high redshift massive galaxies than is seen in local galaxies to
reconcile our PLE models with the data. Note that this dust must be
present {\it without} an accompanying population of younger stars,
which would raise the intrinsic luminosity of the stellar population
of the galaxy. In the nearby universe such populations are almost
always present in dusty galaxies and the effects of the young stars
cancel almost exactly those of the dust, resulting in a
colour--apparent M/L relation which depends very weakly on dust
content \citep{2001ApJ...550..212B,2003MNRAS.341...33K}. If high
redshift galaxies behave similarly, then dust will not help reconcile
our PLE models with the data.

Our main conclusion is that traditional PLE models cannot reconcile the
relatively small number of high redshift galaxies found in deep K-selected
redshift surveys with the abundance of massive galaxies seen in the local
Universe. The counterparts of nearby luminous red galaxies just do not seem to
be present in sufficient numbers at redshifts of 1.5 to 2. The areas of the
deep surveys are quite small, so there may still be significant uncertainties
in this statement as a result of cosmic variance. Substantial amounts of dust
might also cause many distant massive galaxies to be missed, but only if dust
attenuation is not compensated by emission from young stars in the way
observed in low redshift galaxies. If these two possibilities are insufficient
to explain the discrepancy, then one will be forced to conclude that most
nearby massive galaxies were assembled at $z<1.5$, presumably by mergers of
pre-existing stellar systems, since their stars appear to be old.

\bibliographystyle{aa}
\bibliography{references}

\begin{thebibliography}{22}
\expandafter\ifx\csname natexlab\endcsname\relax\def\natexlab#1{#1}\fi

\bibitem[{{Bell} \& {de Jong}(2001)}]{2001ApJ...550..212B}
{Bell}, E.~F. \& {de Jong}, R.~S. 2001, \apj, 550, 212

\bibitem[{{Blanton} {et~al.}(2003){Blanton}, {Hogg}, {Bahcall}, {Baldry},
  {Brinkmann}, {Csabai}, {Eisenstein}, {Fukugita}, {Gunn}, {Ivezi{\' c}},
  {Lamb}, {Lupton}, {Loveday}, \& {Munn}}]{2003ApJ...594..186B}
{Blanton}, M.~R., {Hogg}, D.~W., {Bahcall}, N.~A., {et~al.} 2003, \apj, 594,
  186

\bibitem[{{Bruzual} \& {Charlot}(2003)}]{2003MNRAS.344.1000B}
{Bruzual}, G. \& {Charlot}, S. 2003, \mnras, 344, 1000

\bibitem[{{Cimatti} {et~al.}(2002{\natexlab{a}}){Cimatti}, {Mignoli}, {Daddi},
  {Pozzetti}, {Fontana}, {Saracco}, {Poli}, {Renzini}, {Zamorani},
  {Broadhurst}, {Cristiani}, {D'Odorico}, {Giallongo}, {Gilmozzi}, \&
  {Menci}}]{2002A&A...392..395C}
{Cimatti}, A., {Mignoli}, M., {Daddi}, E., {et~al.} 2002{\natexlab{a}}, \aap,
  392, 395

\bibitem[{{Cimatti} {et~al.}(2002{\natexlab{b}}){Cimatti}, {Pozzetti},
  {Mignoli}, {Daddi}, {Menci}, {Poli}, {Fontana}, {Renzini}, {Zamorani},
  {Broadhurst}, {Cristiani}, {D'Odorico}, {Giallongo}, \&
  {Gilmozzi}}]{2002A&A...391L...1C}
{Cimatti}, A., {Pozzetti}, L., {Mignoli}, M., {et~al.} 2002{\natexlab{b}},
  \aap, 391, L1

\bibitem[{{Firth} {et~al.}(2002){Firth}, {Somerville}, {McMahon}, {Lahav},
  {Ellis}, {Sabbey}, {McCarthy}, {Chen}, {Marzke}, {Wilson}, {Abraham},
  {Beckett}, {Carlberg}, \& {Lewis}}]{2002MNRAS.332..617F}
{Firth}, A.~E., {Somerville}, R.~S., {McMahon}, R.~G., {et~al.} 2002, \mnras,
  332, 617

\bibitem[{{Fontana} {et~al.}(1999){Fontana}, {Menci}, {D'Odorico}, {Giallongo},
  {Poli}, {Cristiani}, {Moorwood}, \& {Saracco}}]{1999MNRAS.310L..27F}
{Fontana}, A., {Menci}, N., {D'Odorico}, S., {et~al.} 1999, \mnras, 310, L27

\bibitem[{{Gardner}(1998)}]{1998PASP..110..291G}
{Gardner}, J.~P. 1998, \pasp, 110, 291

\bibitem[{{Gavazzi} {et~al.}(1991){Gavazzi}, {Boselli}, \&
  {Kennicutt}}]{1991AJ....101.1207G}
{Gavazzi}, G., {Boselli}, A., \& {Kennicutt}, R. 1991, \aj, 101, 1207

\bibitem[{{Kashikawa} {et~al.}(2003){Kashikawa}, {Takata}, {Ohyama}, {Yoshida},
  {Maihara}, {Iwamuro}, {Motohara}, {Totani}, {Nagashima}, {Shimasaku},
  {Furusawa}, {Ouchi}, {Yagi}, {Okamura}, {Iye}, \&
  {Sasaki}}]{2003AJ....125...53K}
{Kashikawa}, N., {Takata}, T., {Ohyama}, Y., {et~al.} 2003, \aj, 125, 53

\bibitem[{{Kauffmann} \& {Charlot}(1998)}]{1998MNRAS.297L..23K}
{Kauffmann}, G. \& {Charlot}, S. 1998, \mnras, 297, L23+

\bibitem[{{Kauffmann} {et~al.}(2003){Kauffmann}, {Heckman}, {White}, {Charlot},
  {Tremonti}, {Brinchmann}, {Bruzual}, {Peng}, {Seibert}, {Bernardi},
  {Blanton}, {Brinkmann}, {Castander}, \& {Cs{\' a}bai}}]{2003MNRAS.341...33K}
{Kauffmann}, G., {Heckman}, T.~M., {White}, S.~D.~M., {et~al.} 2003, \mnras,
  341, 33

\bibitem[{{Kochanek} {et~al.}(2001){Kochanek}, {Pahre}, {Falco}, {Huchra},
  {Mader}, {Jarrett}, {Chester}, {Cutri}, \& {Schneider}}]{2001ApJ...560..566K}
{Kochanek}, C.~S., {Pahre}, M., {Falco}, E., {et~al.} 2001, \apj, 560, 566

\bibitem[{{Mobasher} {et~al.}(2004){Mobasher}, {Idzi}, {Ben{\'{\i}}tez},
  {Cimatti}, {Cristiani}, {Daddi}, {Dahlen}, {Dickinson}, {Erben}, {Ferguson},
  {Giavalisco}, {Grogin}, {Koekemoer}, {Mignoli}, \&
  {Moustakas}}]{2004ApJ...600L.167M}
{Mobasher}, B., {Idzi}, R., {Ben{\'{\i}}tez}, N., {et~al.} 2004, \apjl, 600,
  L167

\bibitem[{{Nagashima} {et~al.}(2004){Nagashima}, {Lacey}, {Baugh}, {Frenk}, \&
  {Cole}}]{2004astro.ph..8529N}
{Nagashima}, M., {Lacey}, C.~G., {Baugh}, C.~M., {Frenk}, C.~S., \& {Cole}, S.
  2004, ArXiv Astrophysics e-prints

\bibitem[{{Rudnick} {et~al.}(2001){Rudnick}, {Franx}, {Rix}, {Moorwood},
  {Kuijken}, {van Starkenburg}, {van der Werf}, {R{\" o}ttgering}, {van
  Dokkum}, \& {Labb{\' e}}}]{2001AJ....122.2205R}
{Rudnick}, G., {Franx}, M., {Rix}, H., {et~al.} 2001, \aj, 122, 2205

\bibitem[{{Somerville} {et~al.}(2004){Somerville}, {Moustakas}, {Mobasher},
  {Gardner}, {Cimatti}, {Conselice}, {Daddi}, {Dahlen}, {Dickinson},
  {Eisenhardt}, {Lotz}, {Papovich}, {Renzini}, \&
  {Stern}}]{2004ApJ...600L.135S}
{Somerville}, R.~S., {Moustakas}, L.~A., {Mobasher}, B., {et~al.} 2004, \apjl,
  600, L135

\bibitem[{{Tinsley}(1980)}]{1980ApJ...241...41T}
{Tinsley}, B.~M. 1980, \apj, 241, 41

\bibitem[{{Totani} {et~al.}(2001){Totani}, {Yoshii}, {Maihara}, {Iwamuro}, \&
  {Motohara}}]{2001ApJ...559..592T}
{Totani}, T., {Yoshii}, Y., {Maihara}, T., {Iwamuro}, F., \& {Motohara}, K.
  2001, \apj, 559, 592

\bibitem[{{van Dokkum} {et~al.}(1999){van Dokkum}, {Franx}, {Fabricant},
  {Kelson}, \& {Illingworth}}]{1999ApJ...520L..95V}
{van Dokkum}, P.~G., {Franx}, M., {Fabricant}, D., {Kelson}, D.~D., \&
  {Illingworth}, G.~D. 1999, \apjl, 520, L95

\bibitem[{{van Dokkum} \& {Stanford}(2003)}]{2003ApJ...585...78V}
{van Dokkum}, P.~G. \& {Stanford}, S.~A. 2003, \apj, 585, 78

\bibitem[{{York} {et~al.}(2000){York}, {Adelman}, {Anderson}, {Anderson},
  {Annis}, {Bahcall}, {Bakken}, {Barkhouser}, {Bastian}, {Berman}, {Boroski},
  {Bracker}, {Briegel}, {Briggs}, {Brinkmann}, \&
  {Brunner}}]{2000AJ....120.1579Y}
{York}, D.~G., {Adelman}, J., {Anderson}, J., {et~al.} 2000, \aj, 120, 1579

\end{thebibliography}

\end{document}